 \documentclass[final,3p,times,twocolumn]{elsarticle}
\usepackage{caption}
\usepackage{subcaption}
\usepackage{amssymb}
\usepackage{amsmath}
\usepackage{svg}
\usepackage{blindtext}
\usepackage{lineno}
\usepackage{hyperref}
\usepackage{pdflscape}
\usepackage{comment}
\usepackage{booktabs}
\usepackage{graphicx}
\usepackage{xcolor}
\usepackage{epstopdf}
\hypersetup{
    colorlinks=true,
    linkcolor=blue,
    filecolor=magenta,      
    urlcolor=cyan,
}

\journal{Bioresource Technology}

\begin{document}

\begin{frontmatter}

%% Title, authors and addresses

\title{Sensor-Based Turbidostat Operation Enables Biomass Setpoint Regulation and Productivity Improvement in semi-industrial Microalgae Raceway Ponds}

\author[1]{José González-Hernández*}
\author[2]{Laura Bernacchioni}
\author[3]{Ainoa Morillas-España}
\author[1]{José Luis Guzmán}
\author[3]{Francisco Gabriel Acién}

%% \author[label1,label2]{<author name>}
\address[1]{University of Almería, Department of Informatics, CIESOL, ceiA3, La Cañada de San Urbano s/n, Almería, Spain. }
\address[2]{University of Bologna, Department of Civil, Chemical, Environmental and Materials Engineering, Via Terracini 28, 40131 Bologna, Italy.}
\address[3]{University of Almería, Chemical Engineering, CIESOL, ceiA3, La Cañada de San Urbano s/n, Almería, Spain.\\
*Corresponding author: José González-Hernández (j.gonzalez@ual.es)}

\begin{abstract}
This work presents the experimental validation of a turbidostat strategy for biomass control in a semi-industrial outdoor raceway reactor. The proposed approach regulates biomass concentration by automatically triggering dilution when the online biomass estimate exceeds a predefined threshold. To ensure safe outdoor operation, dilution was restricted to daylight periods, avoiding biomass removal under low-radiation conditions. The strategy was implemented through an industrial control architecture using an optical monitoring system for online biomass estimation. Experiments were conducted over 14 consecutive days in an 80~m$^2$ (12000~L) raceway reactor. A second parallel reactor operated in chemostat mode, with a nominal dilution of 20\% of the total volume during operating days, provided contextual information under the same outdoor conditions. The analysis focuses on the ability of the sensor-based strategy to configure and maintain the desired biomass concentration, rather than on a direct reactor-to-reactor performance ranking.
During the campaign, the biomass threshold in the turbidostat reactor was changed from 1.0 to 0.8~g~L$^{-1}$, demonstrating the flexibility enabled by online biomass monitoring. Excluding initial adjustment and transition days, harvested areal productivity increased from 9.52 to 23.20~g~m$^{-2}$~d$^{-1}$ after reducing the operating threshold. The overall biomass balance also showed higher net areal productivity in the turbidostat reactor, reaching 20.34~g~m$^{-2}$~d$^{-1}$ compared with 11.16~g~m$^{-2}$~d$^{-1}$ in the parallel chemostat reactor. These results demonstrate the feasibility of robust turbidostat-based biomass control in large-scale outdoor raceway photobioreactors.
\end{abstract}


\end{frontmatter}

%\linenumbers

\section{Introduction}

Microalgae cultivation has gained increasing attention as a sustainable technology for biomass production, wastewater treatment, and CO$_2$ biofixation. Their photosynthetic metabolism enables the conversion of solar radiation and inorganic carbon into biomass, making microalgae suitable for applications within circular economy and carbon mitigation frameworks \citep{Benemann2003,Chisti2007,Acien2017}. In this context, significant progress has been made in the design, modelling, and operation of photobioreactors, supporting the development of cultivation systems at pilot and industrial scales \citep{Acien2017,Guzman2025tutorial}.

Among the available cultivation technologies, open raceway photobioreactors remain one of the most widely used configurations for large-scale microalgae production because of their relatively low capital and operating costs, simple construction, and scalability \citep{Weissman1988,Chisti2007}. However, outdoor raceway operation is strongly affected by environmental disturbances, mainly solar radiation and temperature, which introduce highly time-varying growth dynamics \citep{Bernard2011}. Moreover, biomass concentration is a key variable for productivity and harvesting decisions, but it is still difficult to monitor reliably online in large-scale outdoor systems, limiting the implementation of closed-loop control strategies \citep{Dochain2000,GarciaManas2019}.

In practical large-scale operation, microalgae cultures are frequently managed under batch or chemostat regimes, where harvesting or dilution is applied according to predefined operating criteria or operator experience \citep{MolinaGrima1999}. Although these strategies are simple and robust, they do not necessarily adapt the biomass concentration to the actual growth dynamics of the culture, which may lead to suboptimal operating conditions under variable outdoor environments. Turbidostat operation offers an alternative approach by regulating biomass concentration through dilution, maintaining the culture within a desired optical density or biomass range \citep{Hoskisson2005,deVree2016}.

Despite its potential, turbidostat operation has rarely been demonstrated in large-scale outdoor raceway reactors. Its practical implementation requires not only online information on the culture state, but also reliable actuation and supervision layers capable of operating under variable environmental conditions. Previous works have addressed biomass monitoring using optical measurements or observer-based approaches \citep{Dochain2000,GarciaManas2019}; however, the integration of online biomass information into autonomous harvesting control remains limited in outdoor raceway systems.

Several recent studies have addressed related aspects of outdoor microalgae automation, but with different objectives, sensing approaches, and cultivation scales. The AlgaePARC facility was designed to compare outdoor photobioreactor concepts, including raceway, tubular, and flat-panel systems, with online measurements and the possibility of chemostat or turbidostat operation over approximately 25 m$^2$ ground areas \citep{Bosma2014AlgaePARC}. Later, \citet{deVree2016} evaluated turbidostat operation in outdoor pilot-scale photobioreactors using NIR turbidity probes and SCADA-based harvesting to study the influence of biomass concentration on productivity. More recently, \citet{Gao2026Spectroradiometric} implemented automatic dilution in 100 L outdoor raceway ponds using non-contact spectroradiometric monitoring, reporting productivity improvements with both turbidostat and light-driven operation.  However, scaling up outdoor raceway operation to semi-industrial systems introduces additional challenges associated with distributed hydrodynamics, spatial heterogeneity, and local mass balances across different reactor zones (e.g., sump, circulation channels, and mixing areas), which may affect biomass distribution and control performance. In contrast, the present work focuses on a semi-industrial 80 m$^2$ (12000 L) open raceway reactor and combines a custom flow-through low-cost optical biomass sensor with an industrial control architecture based on an OPC UA communication and a PLC (Programmable Logic Controller).

In outdoor cultivation, turbidostat operation must account for the strong dependence of biomass growth on the daily light cycle. Unlike laboratory systems, where illumination and growth conditions can be kept nearly constant, outdoor raceways are exposed to periods of low or zero irradiance in which dilution may not contribute to biomass production. Therefore, practical turbidostat implementations in outdoor systems should include operational constraints that prevent unnecessary harvesting during unfavorable growth periods \citep{deVree2016}.

In this context, this work presents the experimental validation of a turbidostat strategy for biomass control under real outdoor conditions for 14 consecutive days. The strategy uses online biomass information to activate dilution according to predefined biomass thresholds, while harvesting is enabled only during daylight periods for safe and efficient operation. The main contributions of this work are threefold. First, it provides experimental evidence for turbidostat-based biomass control in a semi-industrial scale outdoor raceway reactor. Second, it demonstrates the integration of online biomass information with an industrial supervision and actuation architecture for autonomous harvesting. Third, the results show that the biomass operating concentration in the turbidostat-based reactor can be intentionally adjusted during outdoor operation, as illustrated by the change from 1.0 to 0.8 g\,L$^{-1}$. In addition, the productivity of a parallel chemostat-based reactor operated with a nominal dilution equivalent to 20\% of the total volume (2400\,L) during operating days is reported as contextual operational information. Although not intended as a strict direct benchmark, the comparison indicates improved productivity under turbidostat operation relative to the chemostat-based strategy.

This configurable operation is also relevant for the future integration of optimization-based supervisory layers. Recent advances in Economic Model Predictive Control (EMPC) for microalgae cultivation systems provide a promising framework for dynamically adapting biomass references according to forecasted environmental conditions, solar radiation availability, and productivity objectives \citep{Otalora2025EMPC}. However, these strategies require reliable online biomass measurements and robust low-level actuation. Therefore, the sensing and turbidostat framework proposed here can be understood as an enabling layer for future predictive and economic optimization approaches.

The remainder of this paper is organized as follows. Section~2 describes the experimental setup, the optical biomass monitoring system, the biomass estimation model, and the implementation of the constrained turbidostat strategy. Section~3 presents and discusses the experimental results, including the biomass estimation performance, the operation of the turbidostat controller under outdoor conditions, the biomass-setpoint change in RW5, and the contextual productivity obtained in the parallel chemostat reactor. Finally, Section~4 summarizes the main conclusions and outlines future research directions.

\section{Materials and Methods}

\subsection{Experimental setup}
\begin{figure}[h]
    \centering
    \includegraphics[width=\linewidth]{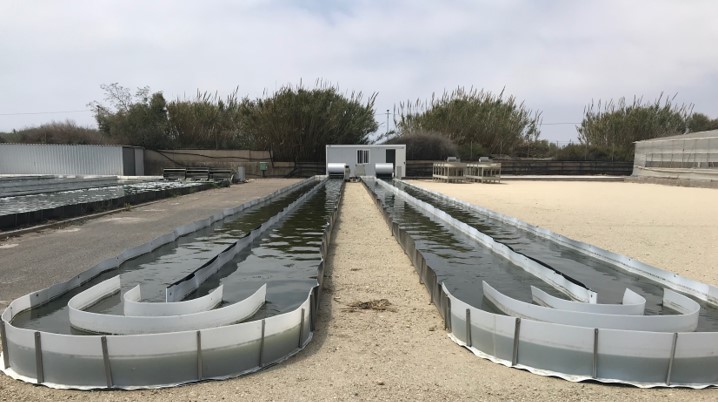}
    \caption{Outdoor 80 m$^2$ raceway photobioreactors (RW5 left and RW6 right) at the IFAPA experimental facility used in this study.}
    \label{fig:Raceway_IFAPA}
\end{figure}

The experiments were carried out in two geometrically identical open raceway photobioreactors located at the IFAPA research facilities in Almería, Spain. Both reactors were inoculated with the same microalgal strain, \textit{Scenedesmus almeriensis}. Reactor RW5 was used to validate the sensor-based turbidostat strategy, whereas reactor RW6 was operated as the parallel chemostat reactor and used to report contextual productivity under the same biological and environmental framework. Each reactor has a surface area of approximately 80 m$^2$ and was operated under outdoor conditions. A general view of the experimental facility is shown in Fig.~\ref{fig:Raceway_IFAPA}.

Each raceway consists of a closed-loop channel formed by two straight sections, approximately 40 m long and 1 m wide, connected by U-shaped bends. Culture circulation is provided by a paddle wheel, ensuring continuous mixing and promoting light-dark cycling along the reactor channel.

The reactors can typically be operated at shallow culture depths between 10 cm and 15 cm; in this work, the culture depth was fixed at 15 cm, corresponding to the optimal operating height reported for this type of raceway photobioreactor~\citep{Gonzalez2022temperatura}. This configuration resulted in a working volume of approximately 12 m$^3$ for each 80 m$^2$ reactor. Both systems were equipped with industrial instrumentation, including pH probes, dissolved oxygen (DO) sensors, temperature sensors, and gas flow meters for CO$_2$ and air injection. CO$_2$ was injected in the sump section for pH regulation and inorganic carbon supply, whereas air injection was used to promote oxygen removal from the culture.

The two reactors were operated in parallel and exposed to the same outdoor environmental conditions. RW6 was used to provide contextual information on the productivity obtained in the second reactor under chemostat operation, while the main analysis focused on the closed-loop biomass regulation achieved in RW5.

\subsection{Instrumentation and data acquisition}

All process variables were collected through the industrial control system with a sampling time of 10 s. The recorded variables included environmental measurements, such as solar radiation and ambient temperature, and process variables, including pH, dissolved oxygen, culture temperature, and CO$_2$ and air flow rates.

Real-time communication between the monitoring devices, supervisory computer, and PLC was implemented using an OPC UA architecture. This communication layer enabled online data exchange for process supervision and control implementation.

\subsection{Optical sensing system for biomass monitoring}

Online biomass information was provided by an optical sensing system implemented to support turbidostat operation in the outdoor raceway reactor. Since biomass concentration is not available from the standard industrial instrumentation of the raceway, this module provided the real-time estimate required to detect biomass threshold crossings and activate harvesting decisions.

The sensing system is based on two multichannel spectral sensors (AS7341, AMS-OSRAM) connected to a Raspberry Pi 4 via I2C. Measurements are performed in a flow-through cuvette where culture samples are periodically introduced from the reactor. Two complementary optical configurations are used:

\begin{itemize}
    \item \textbf{Absorbance measurement:} a white LED light source is aligned with one spectral sensor to measure transmitted light at multiple wavelengths.
    \item \textbf{Fluorescence measurement:} a second spectral sensor is positioned at 90$^\circ$ with respect to a 450 nm excitation source to measure fluorescence emission while reducing the influence of direct excitation light.
\end{itemize}

The measurement sequence is fully automated and includes blank measurement using clean water, sample acquisition from the reactor, absorbance measurement, fluorescence measurement, and flushing of the hydraulic circuit. This sequence reduces the influence of fouling and baseline drift, which are common limitations in optical measurements under outdoor cultivation conditions.

The measurement cycle is executed every 5 min, with a total duration of approximately 120 s. Data acquisition and signal processing are implemented in Python on the embedded platform. The processed variables are stored locally and exposed through an OPC UA server, allowing the biomass estimate to be accessed by the supervisory control layer.

\begin{figure*}[h]
\centering
\includegraphics[width=\linewidth]{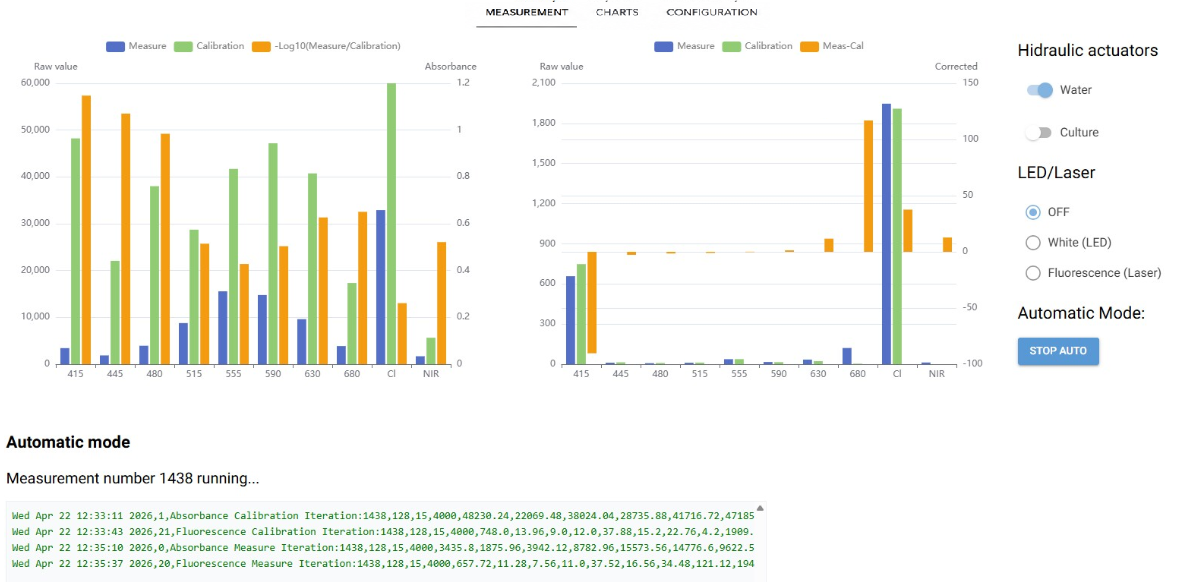}
\caption{Interface for real-time visualization and configuration of the optical biomass monitoring system.}
\label{fig:interface}
\end{figure*}

A web-based interface was also developed for real-time visualization and configuration of the optical sensing system. The interface allows the visualization of spectral measurements, calibration signals, and processed variables, as shown in Fig.~\ref{fig:interface}.

\begin{figure}[h!]
\centering
\begin{subfigure}{0.48\linewidth}
    \centering
    \includegraphics[width=\linewidth]{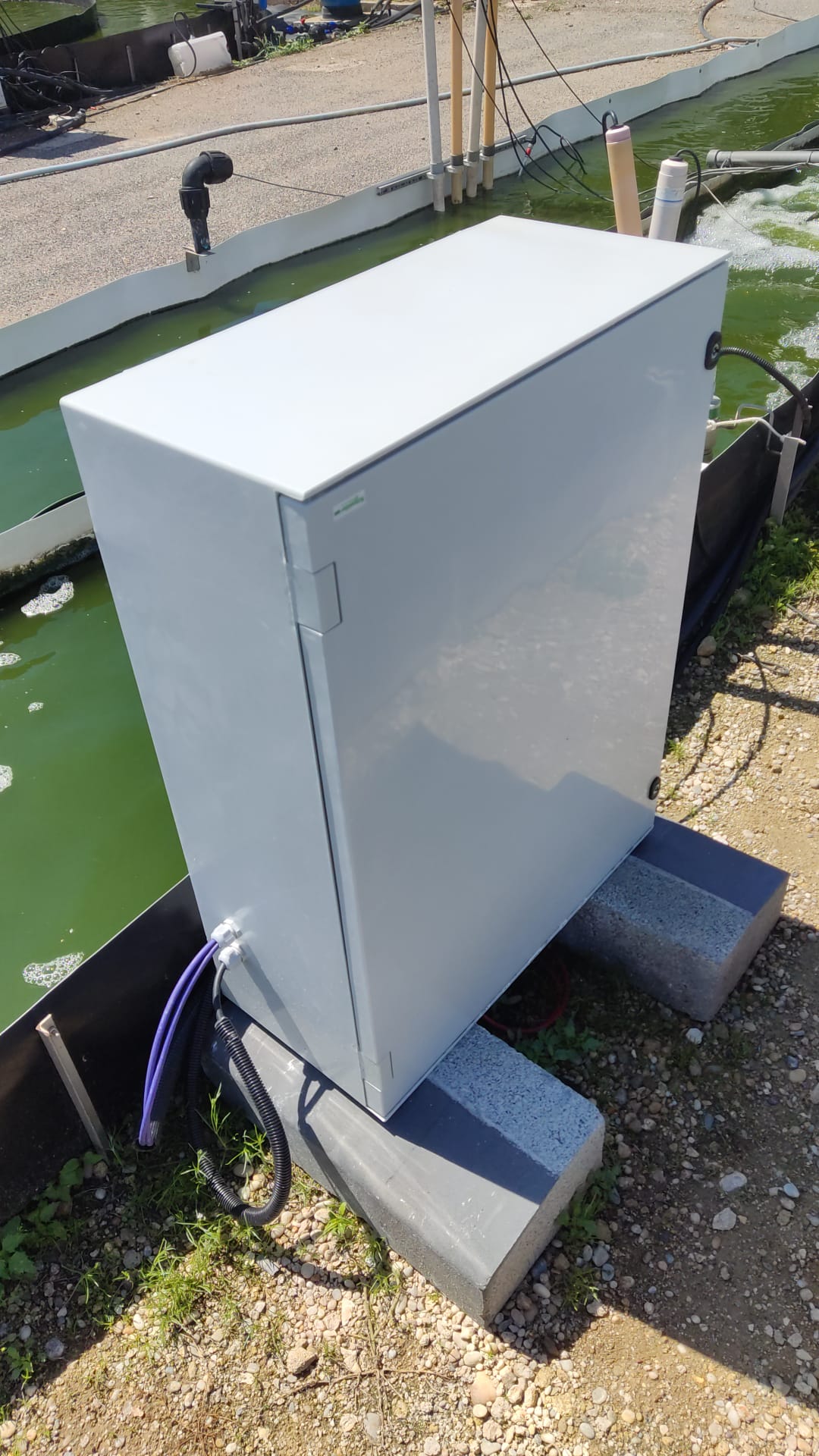}
    \caption{External view of the sensing system installed next to the raceway reactor.}
\end{subfigure}
\hfill
\begin{subfigure}{0.48\linewidth}
    \centering
    \includegraphics[width=\linewidth]{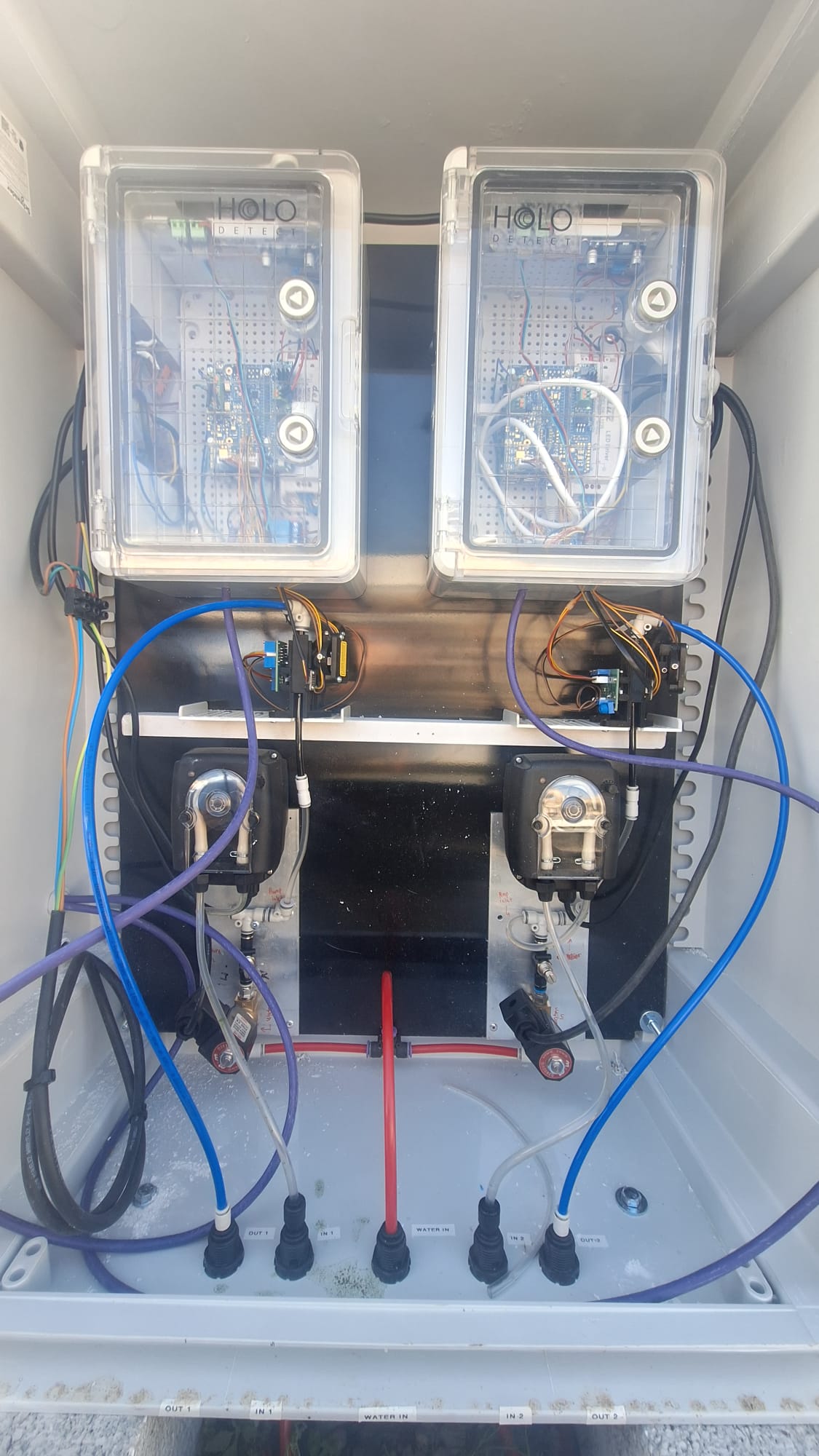}
    \caption{Internal view of the sensing system showing pumps, electronics, and flow-through measurement setup.}
\end{subfigure}
\caption{Optical sensing system used to provide online biomass information for turbidostat operation.}
\label{fig:sensor_system}
\end{figure}

The sensing system was installed in a protective enclosure located next to the raceway reactor, enabling continuous outdoor operation during the experimental campaign. Figure~\ref{fig:sensor_system} shows the external deployment and internal configuration of the device.

\subsection{Biomass estimation model}

Biomass concentration was estimated from the optical measurements using a data-driven calibration model. The calibration dataset was obtained by pairing absorbance and fluorescence signals measured at multiple wavelengths with offline dry weight measurements used as reference biomass concentrations.

A linear regression model with L1 regularization, commonly referred to as LASSO, was used:

\begin{equation}
\hat{X}(t) = \beta_0 + \sum_{i=1}^{n} \beta_i \, s_i(t),
\end{equation}
where $\hat{X}(t)$ is the estimated biomass concentration, $s_i(t)$ denotes the $i$-th optical feature obtained from the sensing system, $\beta_0$ is the intercept, and $\beta_i$ are the regression coefficients.

The LASSO regularization promotes sparse coefficient vectors, allowing the model to select the most informative optical features and reducing the risk of overfitting \citep{Tibshirani1996}. During the experiments, the resulting model provided the online biomass estimate used by the turbidostat control algorithm. Its estimation performance was evaluated against independent dry weight measurements, as reported in Section 3.1.

During the experiments, the resulting model provided the online biomass estimate used by the turbidostat control algorithm.

\subsection{Control strategy: constrained turbidostat}

The control objective is to maintain the biomass concentration within a predefined band:

\begin{equation}
X_{\min} \leq \hat{X}(t) \leq X_{\max}
\end{equation}

To avoid chattering and ensure complete dilution cycles, a hysteresis-based on-off control strategy is adopted. The dilution state is defined by a binary variable $u(t) \in \{0,1\}$ governed by:

\begin{equation}
u(t) =
\begin{cases}
1, & \text{if } \hat{X}(t) > X_{\max} \ \text{and} \ t \in \mathcal{T}_{\text{light}} \\
0, & \text{if } \hat{X}(t) < X_{\min} \ \text{or} \ t \notin \mathcal{T}_{\text{light}} \\
u(t^-), & \text{otherwise}
\end{cases}
\end{equation}
where $\mathcal{T}_{\text{light}} = [09{:}00, 20{:}00]$ denotes the daily period in which the algorithm is activated and $u(t^-)$ means to keep the previous $u(t)$ value.

Thus, the dilution flow rate is then given by:
\begin{equation}
Q_d(t) = u(t)\, Q_{\max}
\end{equation}
where $Q_{\max}$ is the maximum dilution rate.

Dilution rate is implemented through level control. When activated, fresh medium is added, causing overflow and biomass removal.

As the control action is implemented in an on-off manner with hysteresis, it provides a simple and robust strategy suitable for industrial environments.

\subsection{Control implementation architecture}

\begin{figure}[h!]
\centering
\includegraphics[width=\linewidth]{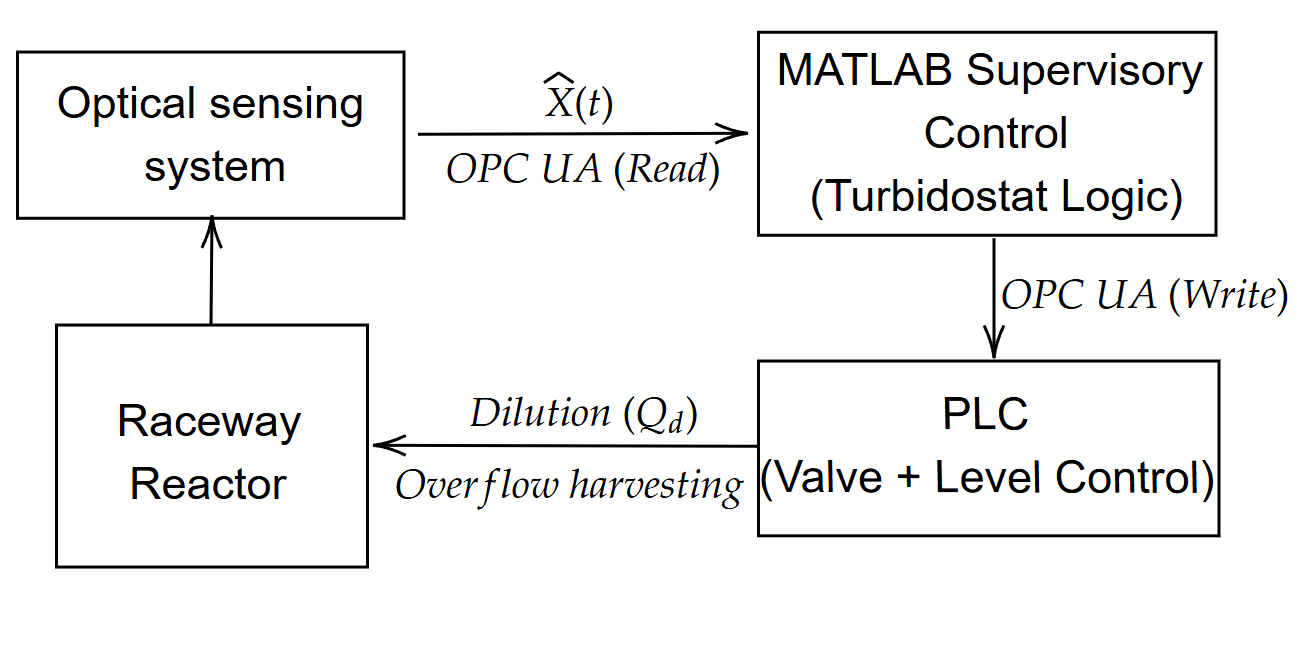}
\caption{Control architecture used for turbidostat implementation, including online biomass estimation, OPC UA communication, MATLAB supervision, and PLC-based dilution and overflow harvesting.}
\label{fig:control_scheme}
\end{figure}

Figure~\ref{fig:control_scheme} illustrates the implemented control architecture. The biomass concentration estimate $\hat{X}(t)$ is obtained from the optical sensing system and published to an OPC UA server. This data is accessed by a supervisory control algorithm implemented in MATLAB, which evaluates the turbidostat condition based on the predefined thresholds.

When the control condition is satisfied, the MATLAB-based supervisory layer sends a command via OPC UA to the PLC, which acts as the actuator layer. The PLC is responsible for opening or closing the dilution valve, thereby injecting fresh medium into the raceway reactor.

Dilution is physically implemented through level control. The PLC internally regulates the liquid level by harvesting culture when fresh medium is added, ensuring a constant reactor height and volume. As a result, biomass is removed via overflow while maintaining stable operating conditions.

This architecture combines high-level decision making in MATLAB with reliable low-level execution in the PLC, providing a robust and industry-compatible turbidostat implementation.

\subsection{Experimental protocol and evaluation metrics}

As commented above, the experimental campaign was conducted using two geometrically identical raceway photobioreactors operated in parallel under the same outdoor environmental conditions. Reactor RW5 was operated using the proposed turbidostat strategy, whereas reactor RW6 was operated in chemostat mode. This parallel configuration was not intended to establish a strict direct comparison between control strategies, because the reactors were operated under different objectives and biomass trajectories. Instead, RW6 provides the productivity obtained in the second reactor under its own operating mode.

In RW5, harvesting was automatically activated according to the turbidostat logic described above, using the online biomass estimate and the predefined biomass thresholds. In RW6, culture was removed under chemostat operation, with fresh medium addition and culture overflow maintaining approximately constant reactor volume.

Offline dry weight measurements were used as the reference method for biomass concentration. For each sampling time, three culture samples were collected from each reactor. Sampling was performed twice per day, at approximately 08:00 and 14:00. The samples were filtered and dried to constant weight following standard laboratory procedures. For each sampling time, the mean biomass concentration and standard deviation were calculated from the three replicates. These values were used to evaluate the biomass concentration evolution and to compare the online biomass estimates with the reference measurements.

The performance of both operating strategies was evaluated using the following indicators:

\begin{itemize}
    \item Biomass concentration.
    \item Harvested culture volume.
    \item Cumulative harvested biomass.
\end{itemize}

The harvested biomass was calculated differently for each operating strategy. In RW5, harvesting was activated by the turbidostat controller over short dilution events; therefore, the biomass concentration during each harvested volume was assumed to be approximately constant and equal to the online biomass concentration at the time of harvesting. The cumulative harvested biomass in RW5 was then calculated as the sum of the biomass removed during all harvesting events.

In RW6, harvesting was performed as a chemostat operation at approximately constant reactor volume, using a nominal daily dilution of 20\% of the total reactor volume, corresponding to 2400 L during the operating days. In this case, fresh medium addition and culture overflow occurred simultaneously, so the biomass concentration of the harvested stream decreased during the harvesting period. Therefore, the biomass removed in RW6 was estimated from the biomass concentrations before and after harvesting, accounting for the variation in outlet concentration during the process. The cumulative harvested biomass was obtained as the sum of the biomass removed over the experimental campaign.

\section{Results and Discussion}
This section presents the experimental validation of the proposed sensing and control framework under real outdoor conditions for 14 consecutive days of operation. The performance of the online biomass estimation system is evaluated against offline dry weight measurements. Subsequently, the behavior of the turbidostat strategy is analyzed in terms of biomass regulation, configurable biomass operation, and harvested biomass production. The RW6 chemostat run is reported as contextual information from the parallel reactor, avoiding a direct one-to-one performance comparison.

\subsection{Biomass estimation performance}

\begin{landscape}
\begin{figure}[p]
    \centering
    \includegraphics[width=1.15\linewidth]{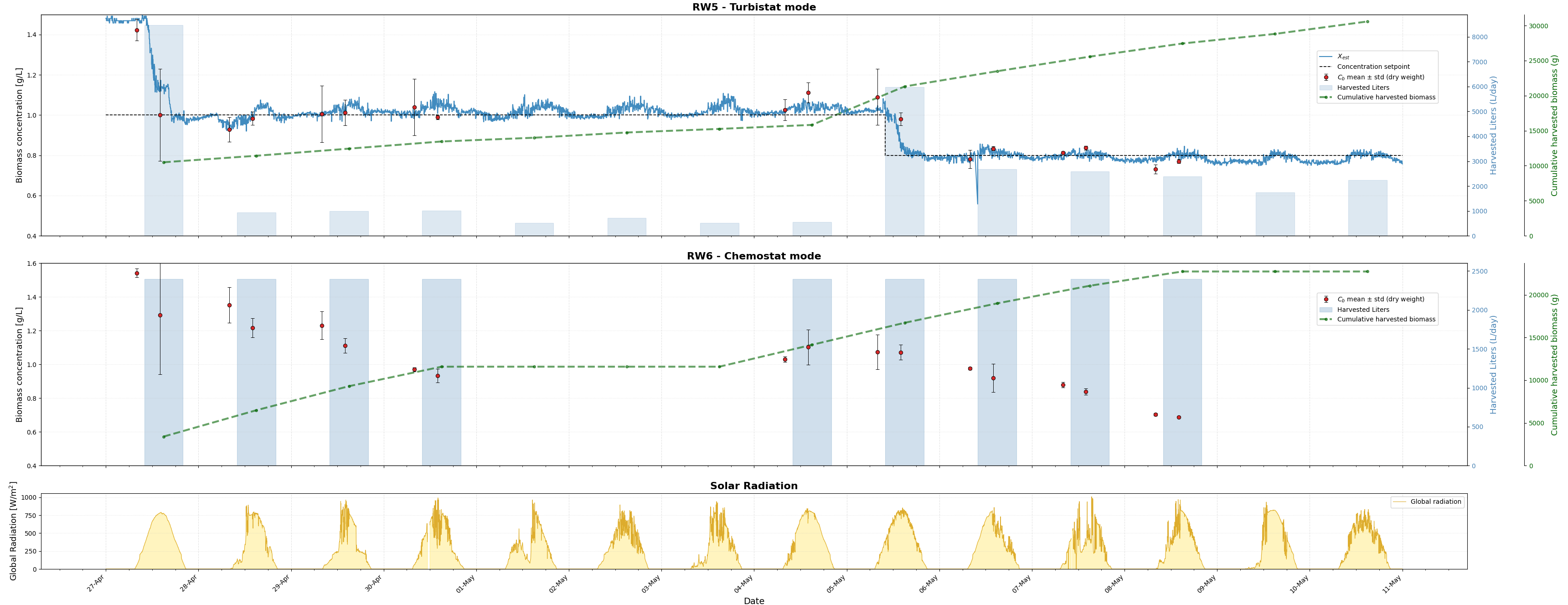}
    \caption{Experimental results obtained during outdoor operation of the two raceway photobioreactors. The figure shows the online biomass estimation $\hat{X}(t)$, offline dry weight measurements ($C_{b,\mathrm{mean}} \pm \sigma$), harvested culture volume, cumulative harvested biomass, and incident solar radiation. RW5 operated under the proposed constrained turbidostat strategy, with the biomass threshold changed from 1.0 to 0.8~g~L$^{-1}$ during the campaign. RW6 was operated in chemostat mode with a nominal 20\% daily dilution of the total volume (2400~L) during operating days and is reported as contextual information from the parallel reactor. The RW6 biomass trajectory shows a marked decrease during consecutive harvesting days and partial recovery during weekend batch periods, indicating that the imposed 20\% dilution was not compatible with maintaining a biomass concentration close to 1.0~g~L$^{-1}$ under the tested conditions.}
    \label{fig:turbidostat_results}
\end{figure}
\end{landscape}

The proposed optical sensing system was evaluated during outdoor operation by comparing the online biomass estimates $\hat{X}(t)$ with offline dry weight measurements used as reference values. The top plot of Figure~\ref{fig:turbidostat_results} shows the evolution of the estimated biomass concentration together with the corresponding offline measurements for both raceway reactors.

The estimator was able to accurately reproduce the biomass dynamics under varying environmental conditions, including significant fluctuations in solar radiation and continuous outdoor operation over several days. Despite the presence of measurement noise and the challenging operating conditions associated with outdoor cultivation, the estimated biomass concentration remained consistent with the offline dry weight measurements throughout the experimental campaign.

A quantitative evaluation performed over 18 paired samples yielded a mean absolute error (MAE) of 0.042 g/L and a root mean square error (RMSE) of 0.052 g/L. These results indicate that the proposed sensing system provides sufficient accuracy for real-time control applications in outdoor raceway photobioreactors.

The obtained estimation performance proved adequate for closed-loop operation. In particular, the online estimation reliably detected biomass threshold crossings required by the hysteresis-based turbidostat strategy, enabling stable activation and deactivation of the dilution process without oscillatory behavior or excessive switching.

Overall, the results demonstrate the feasibility of integrating low-cost optical instrumentation with industrial communication architectures for continuous biomass monitoring and autonomous operation of large-scale outdoor cultivation systems.

\subsection{Turbidostat performance under outdoor conditions}

Figure~\ref{fig:turbidostat_results} (top plot) presents the experimental results obtained during the operation of reactor RW5 under the proposed constrained turbidostat strategy. The online biomass estimation $\hat{X}(t)$ remained within the desired operating region during most of the experimental campaign despite the strong variability in solar radiation observed under outdoor conditions.

The experiment started with the biomass operating threshold fixed at 1.0~g~L$^{-1}$. After the initial adjustment day, RW5 was operated around this threshold from 2026-04-28 to 2026-05-04. During this period, the reactor harvested 5.33~kg of biomass, corresponding to a harvested areal productivity of 9.52~g~m$^{-2}$~d$^{-1}$. The biomass threshold was then deliberately reduced to 0.8~g~L$^{-1}$, producing a transition harvest on 2026-05-05. This transition day was reported separately and was not considered representative of steady operation at the new biomass threshold. During the subsequent operation around 0.8~g~L$^{-1}$, from 2026-05-06 to 2026-05-10, RW5 harvested 9.28~kg of biomass and reached a harvested areal productivity of 23.20~g~m$^{-2}$~d$^{-1}$. This result indicates that, for this campaign, the lower biomass concentration improved productive performance, likely due to reduced self-shading and better light availability inside the culture. This behavior is consistent with previous studies reporting the existence of an optimal biomass concentration range in raceway systems resulting from the trade-off between light utilization and self-shading effects \citep{Bernard2011}.

The offline dry weight measurements also exhibited relatively low dispersion around the estimated biomass trajectory throughout the experiment, indicating that the proposed sensing and control framework remained stable during long-term outdoor operation. Although some deviations between online estimation and offline measurements were observed, the estimation accuracy proved sufficient for reliable closed-loop operation and automatic harvesting activation.

The cumulative harvested biomass, represented by the green dashed line in Fig.~\ref{fig:turbidostat_results}, showed a continuous increase during the experimental campaign, confirming sustained biomass production under the proposed operating strategy despite the highly variable environmental conditions.

\subsection{RW6 chemostat operation as contextual reference}

Figure~\ref{fig:turbidostat_results} also includes the biomass evolution and harvested biomass obtained in RW6, which was operated in chemostat mode with a nominal 20\% daily dilution of the total reactor volume, corresponding to 2400 L during the operating days. These data are reported to document the behavior of the second reactor under the same outdoor campaign, but they should not be interpreted as a strict direct benchmark against RW5. The two reactors followed different operating objectives: RW5 was controlled to maintain a configurable biomass concentration, whereas RW6 provided the productivity associated with the selected chemostat operation.

Under these conditions, RW6 harvested 22.72~kg of biomass over the campaign, corresponding to a campaign-averaged harvested areal productivity of 20.29~g~m$^{-2}$~d$^{-1}$. However, the biomass concentration in RW6 decreased substantially during the consecutive chemostat harvesting days, reaching a final concentration of 0.69~g~L$^{-1}$. The temporary recovery observed during the weekend, when RW6 remained in batch operation because no chemostat dilution was applied, reinforces this interpretation. Therefore, under the tested outdoor conditions, maintaining a biomass concentration close to 1.0~g~L$^{-1}$ was not compatible with a fixed 20\% daily dilution rate in RW6. This behavior is reported as operational context for the chemostat run and should be considered together with the specific dilution schedule, weekend operation, initial biomass concentration, and reactor-to-reactor variability.

\subsection{Discussion}

The main result is therefore not a direct ranking between RW5 and RW6, but that the sensor-based turbidostat allowed RW5 to impose and modify the biomass operating concentration during the outdoor campaign. In particular, RW5 showed a much lower harvested areal productivity during operation around 1.0~g~L$^{-1}$ than during operation around 0.8~g~L$^{-1}$, increasing from 9.52 to 23.20~g~m$^{-2}$~d$^{-1}$ when the transition day was excluded from both steady operating periods. This supports the idea that, in the tested conditions, RW5 was more productive at 0.8~g~L$^{-1}$ than at 1.0~g~L$^{-1}$, and that the online biomass sensor made it possible to select and maintain that operating concentration.

\begin{table*}[h!]
\centering
\resizebox{\textwidth}{!}{%
\begin{tabular}{lrrrr}
\toprule
Date / Variable &
Harvested volume RW5 (L~d$^{-1}$) &
Harvested biomass RW5 (g~d$^{-1}$) &
Harvested volume RW6 (L~d$^{-1}$) &
Harvested biomass RW6 (g~d$^{-1}$) \\
\midrule
2026-04-27 & 8480.00 & 10489.00 & 2400.00 & 3401.67 \\
2026-04-28 & 940.00 & 928.32 & 2400.00 & 3082.00 \\
2026-04-29 & 1000.00 & 1014.90 & 2400.00 & 2812.00 \\
2026-04-30 & 1010.00 & 1023.33 & 2400.00 & 2284.00 \\
2026-05-01 & 520.00 & 534.85 & 0.00 & \text{--} \\
2026-05-02 & 720.00 & 747.12 & 0.00 & \text{--} \\
2026-05-03 & 510.00 & 512.37 & 0.00 & \text{--} \\
2026-05-04 & 550.00 & 572.60 & 2400.00 & 2562.00 \\
2026-05-05 & 5980.00 & 5483.61 & 2400.00 & 2574.00 \\
2026-05-06 & 2680.00 & 2177.85 & 2400.00 & 2276.60 \\
2026-05-07 & 2590.00 & 2078.11 & 2400.00 & 2061.33 \\
2026-05-08 & 2380.00 & 1876.25 & 2400.00 & 1669.33 \\
2026-05-09 & 1740.00 & 1368.74 & 0.00 & \text{--} \\
2026-05-10 & 2230.00 & 1778.31 & 0.00 & \text{--} \\
\midrule
Total harvested volume (L) & 31330.00 & {} & 21600.00 & {} \\
Total harvested biomass (g) & {} & 30585.38 & {} & 22722.93 \\
\midrule
Initial $C_b$ (g~L$^{-1}$) & {} & 1.42 & {} & 1.54 \\
Final $C_b$ (g~L$^{-1}$) & {} & 0.77 & {} & 0.69 \\
Initial standing biomass (g) & {} & 17040.00 & {} & 18480.00 \\
Final standing biomass (g) & {} & 9240.00 & {} & 8253.33 \\
Harvested + final standing biomass (g) & {} & 39825.38 & {} & 30976.26 \\
Net biomass production (g) & {} & 22785.38 & {} & 12496.26 \\
Net areal productivity (g~m$^{-2}$~d$^{-1}$) & {} & 20.34 & {} & 11.16 \\
\midrule
RW5 operation around 1.0~g~L$^{-1}$ &
\multicolumn{4}{c}{2026-04-28 to 2026-05-04} \\

RW5 harvested biomass around 1.0~g~L$^{-1}$ (g) &
\multicolumn{4}{c}{5333.49} \\

RW5 harvested areal productivity around 1.0~g~L$^{-1}$ (g~m$^{-2}$~d$^{-1}$) &
\multicolumn{4}{c}{9.52} \\

\addlinespace
RW5 setpoint-transition day &
\multicolumn{4}{c}{2026-05-05} \\

RW5 biomass harvested during setpoint transition (g) &
\multicolumn{4}{c}{5483.61} \\

\addlinespace
RW5 operation around 0.8~g~L$^{-1}$ &
\multicolumn{4}{c}{2026-05-06 to 2026-05-10} \\

RW5 harvested biomass around 0.8~g~L$^{-1}$ (g) &
\multicolumn{4}{c}{9279.26} \\

RW5 harvested areal productivity around 0.8~g~L$^{-1}$ (g~m$^{-2}$~d$^{-1}$)&
\multicolumn{4}{c}{23.20} \\

\addlinespace
RW6 campaign-averaged harvested areal productivity under chemostat operation (g~m$^{-2}$~d$^{-1}$) &
\multicolumn{4}{c}{20.29}\\
\bottomrule
\end{tabular}
}
\caption{Daily harvested volume and biomass, together with the overall biomass balance for RW5 and RW6. RW6 is reported as the parallel chemostat reactor and is used as contextual operational information. Net biomass production was calculated as harvested biomass plus final standing biomass minus initial standing biomass. The RW5 productivity at 1.0~g~L$^{-1}$ was calculated after excluding the initial adjustment day, and the productivity at 0.8~g~L$^{-1}$ was calculated after excluding the setpoint-transition day.}
\label{tab:harvested_biomass_balance}
\end{table*}

The overall biomass balance is summarized in Table~\ref{tab:harvested_biomass_balance}. RW5 harvested 30.59~kg and finished with 9.24~kg of standing biomass, while RW6 harvested 22.72~kg and finished with 8.25~kg of standing biomass. When both harvested biomass and final standing biomass were considered, and the initial standing biomass was subtracted, the net biomass production was 22.79~kg in RW5 and 12.50~kg in RW6. This corresponds to net areal productivities of 20.34 and 11.16~g~m$^{-2}$~d$^{-1}$, respectively. Thus, over the complete experimental campaign, the turbidostat reactor achieved a higher mass-balance productivity while maintaining autonomous and configurable biomass regulation. The interpretation remains conservative: RW6 represents the productivity of the parallel chemostat reactor, whereas the central contribution of this work is the autonomous configuration and regulation of biomass concentration in RW5.

This interpretation is consistent with previous evidence that biomass-controlled operation can enhance outdoor productivity, although differences in reactor design, strain, location, scale, and operating protocol must always be considered. In the AlgaePARC turbidostat experiments, \citet{deVree2016} reported that an open raceway pond operated at a lower biomass concentration reached higher areal productivity than operation at a higher biomass concentration, emphasizing the relevance of avoiding excessive self-shading. \citet{Gao2026Spectroradiometric} also reported increased productivity under automated turbidostat operation in 100 liters raceway ponds using non-contact spectroradiometry. The present results extend these observations to a substantially larger 80 m$^2$ raceway (12000 liters) and show that comparable biomass-control principles can be implemented using a low-cost optical sensing unit connected to an industrial OPC UA/PLC control layer.

\section{Conclusions}

This work experimentally validated a constrained turbidostat strategy for real-time biomass control in a large-scale outdoor raceway photobioreactor. The proposed framework integrated a low-cost optical biomass sensing unit, daylight-constrained automatic harvesting, and OPC UA/PLC-based supervision, providing a practical implementation of sensor-based biomass regulation under semi-industrial outdoor conditions.

The online biomass estimator achieved sufficient accuracy for closed-loop operation, with an MAE of 0.042~g~L$^{-1}$ and an RMSE of 0.052~g~L$^{-1}$ with respect to offline dry weight measurements. This enabled reliable detection of biomass threshold crossings and autonomous activation and deactivation of the harvesting process despite variable outdoor conditions.

The results showed that the biomass operating concentration in RW5 could be configured during the campaign, shifting from operation under the 1.0~g~L$^{-1}$ condition to operation under the 0.8~g~L$^{-1}$ condition. Under the tested conditions, the lower biomass concentration led to higher harvested areal productivity, increasing from 9.52 to 23.20~g~m$^{-2}$~d$^{-1}$ after excluding adjustment and transition days.

Over the complete outdoor trial, the turbidostat reactor also achieved a higher net areal productivity than the parallel fixed-dilution chemostat reactor, with 20.34~g~m$^{-2}$~d$^{-1}$ in RW5 compared with 11.16~g~m$^{-2}$~d$^{-1}$ in RW6. These results demonstrate that integrating low-cost optical sensing with industrial automatic control enables stable, configurable, and productive biomass operation in outdoor raceway systems.

\section*{Acknowledgments}
This work has been financed by the following projects: PID2023-150739OB-I00 and PDC2025-165379-I00 financed by the Spanish Ministry of Science and also by the European Union (Grant agreement IDs: 101060991, REALM; 101214199, ALLIANCE).  

\bibliographystyle{elsarticle-harv}
\biboptions{authoryear}
\bibliography{references}

@inproceedings{Benemann2003,
  author = {Benemann, J. R.},
  title = {Biofixation of CO2 and greenhouse gas abatement with microalgae},
  booktitle = {Asia-Pacific Conference on Algal Biotechnology},
  year = {2003}
}

@incollection{Acien2017,
  author = {Acien, F. G. and Fernandez-Sevilla, J. M. and Molina-Grima, E.},
  title = {Microalgae: The basis of mankind sustainability},
  booktitle = {Case Study of Innovative Projects},
  publisher = {InTech},
  year = {2017},
  pages = {123--140}
}

@article{Chisti2007,
  author = {Chisti, Y.},
  title = {Biodiesel from microalgae},
  journal = {Biotechnology Advances},
  volume = {25},
  number = {3},
  pages = {294--306},
  year = {2007}
}

@inproceedings{Guzman2025tutorial,
author = {Guzmán, José Luis and Berenguel, M. and Rodriguez Miranda, Enrique and Acien, Gabriel},
year = {2025},
month = {07},
pages = {4305-4322},
title = {Microalgae production at industrial scale: modelling and control challenges},
doi = {10.23919/ACC63710.2025.11107686}
}

@article{Weissman1988,
  author = {Weissman, J. C. and Goebel, R. P. and Benemann, J. R.},
  title = {Photobioreactor design: mixing, carbon utilization, and oxygen accumulation},
  journal = {Biotechnology and Bioengineering},
  volume = {31},
  pages = {336--344},
  year = {1988}
}

@article{Bernard2011,
  author = {Bernard, O.},
  title = {Hurdles and challenges for modelling and control of microalgae for CO2 mitigation and biofuel production},
  journal = {Journal of Process Control},
  volume = {21},
  pages = {1378--1389},
  year = {2011}
}

@article{Dochain2000,
  author = {Dochain, D.},
  title = {State observers for tubular reactors with unknown kinetics},
  journal = {Journal of Process Control},
  volume = {10},
  pages = {259--268},
  year = {2000}
}

@article{MolinaGrima1999,
  author = {Molina-Grima, E. and Fernandez, J. and Acien, F. G. and Chisti, Y.},
  title = {Photobioreactors: light regime, mass transfer, and scaleup},
  journal = {Journal of Biotechnology},
  volume = {70},
  pages = {231--247},
  year = {1999}
}

@article{Hoskisson2005,
  author = {Hoskisson, P. A. and Hobbs, G.},
  title = {Continuous culture—making a comeback?},
  journal = {Microbiology},
  volume = {151},
  pages = {3153--3159},
  year = {2005}
}

@article{GarciaManas2019,
  author = {García-Mañas, F. and Guzmán, J. L. and Berenguel, M. and Acien, F. G.},
  title = {Biomass estimation of an industrial raceway photobioreactor using an extended Kalman filter},
  journal = {Algal Research},
  volume = {37},
  pages = {103--114},
  year = {2019}
}

@misc{Otalora2025EMPC,
  title         = {Enhancing industrial microalgae production through Economic Model Predictive Control},
  author        = {Ot{\'a}lora, Pablo and Skogestad, Sigurd and Guzm{\'a}n, Jos{\'e} Luis and Berenguel, Manuel},
  year          = {2025},
  eprint        = {2512.15668},
  archivePrefix = {arXiv},
  primaryClass  = {eess.SY},
  doi           = {10.48550/arXiv.2512.15668},
  url           = {https://arxiv.org/abs/2512.15668}
}

@article{deVree2016,
  title={Turbidostat operation of outdoor pilot-scale photobioreactors},
  author={de Vree, J. H. and Bosma, R. and Janssen, M. and Barbosa, M. J. and Wijffels, R. H.},
  journal={Algal Research},
  volume={18},
  pages={198--208},
  year={2016}
}

@article{Bosma2014AlgaePARC,
  author  = {Bosma, R. and de Vree, J. H. and Slegers, P. M. and Janssen, M. and Wijffels, R. H. and Barbosa, M. J.},
  title   = {Design and construction of the microalgal pilot facility {AlgaePARC}},
  journal = {Algal Research},
  volume  = {6},
  pages   = {160--169},
  year    = {2014},
  doi     = {10.1016/j.algal.2014.10.006}
}

@article{Gao2026Spectroradiometric,
  author  = {Gao, Song and Katinas, Christopher and Timlin, Jerilyn A. and Reichardt, Thomas A. and Edmundson, Scott and Huesemann, Michael},
  title   = {Automating microalgal raceway pond operations using real-time spectroradiometric monitoring},
  journal = {Algal Research},
  year    = {2026},
  note    = {In press, article 104706},
  doi     = {10.1016/j.algal.2026.104706}
}

@article{Tibshirani1996,
  author  = {Tibshirani, Robert},
  title   = {Regression Shrinkage and Selection via the Lasso},
  journal = {Journal of the Royal Statistical Society: Series B (Methodological)},
  volume  = {58},
  number  = {1},
  pages   = {267--288},
  year    = {1996}
}

@article{Gonzalez2022temperatura,
author = {González-Hernández, José and Rodriguez Miranda, Enrique and Guzmán, José Luis and Fernández, Francisco and Visioli, Antonio},
year = {2022},
month = {02},
pages = {164-173},
title = {Optimización de temperatura en reactores raceway para la producción de microalgas mediante regulación de nivel},
volume = {19},
journal = {Revista Iberoamericana de Automática e Informática industrial},
doi = {10.4995/riai.2022.16586}
}

\end{document}